\documentclass[pre,floatfix,twocolumn,showpacs]{revtex4}
\usepackage{amsmath}
\usepackage{amsfonts}
\usepackage{epsfig}

\begin{document}
\title{Inverse cubic law of index fluctuation distribution in Indian markets}
\author{Raj Kumar Pan}
\email{rajkp@imsc.res.in}
\author{Sitabhra Sinha}
\email{sitabhra@imsc.res.in}%
%
\affiliation{The Institute of Mathematical Sciences, C.I.T. Campus, Taramani,
Chennai - 600 113 India}
\date{\today}

\begin{abstract}
One of the principal statistical features characterizing the activity in
financial markets is the distribution of fluctuations in market indicators
such as the index. While the developed stock markets, e.g., the New York
Stock Exchange (NYSE) have been found to show heavy-tailed return
distribution with a characteristic power-law exponent, the universality of
such behavior has been debated, particularly in regard to emerging
markets.  Here we investigate the distribution of several indices from the
Indian financial market, one of the largest emerging markets in the world.
We have used tick-by-tick data from the National Stock Exchange (NSE), as
well as, daily closing data from both NSE and Bombay Stock Exchange (BSE).
We find that the cumulative distributions of index returns have long tails
consistent with a power-law having exponent $\alpha \approx 3$, at
time-scales of both 1 min and 1 day.  This ``inverse cubic law'' is
quantitatively similar to what has been observed in developed markets,
thereby providing strong evidence of universality in the behavior of
market fluctuations.
\end{abstract}

\maketitle
\section{Introduction}
Financial markets can be viewed as complex systems with a large number of
interacting components that are subject to external influences or
information flow.  Physicists are being attracted in increasing numbers to
the study of financial markets by the prospect of discovering
universalities in their statistical
properties~\cite{Mantegna99,Bouchaud03,Chatterjee06}.  This has partly been
driven by the availability of large amounts of electronically recorded data
with very high temporal resolution, making it possible to study various
indicators of market activity. Among the various candidates for
market-invariant features, the most widely studied are the distributions of
fluctuations in overall market indicators such as {\em market indices}. 

To study these fluctuations such that the result is independent of the
scale of measurement, we define the logarithmic return for a time
scale $\Delta t$ as, 
\begin{equation}
R(t,\Delta t) \equiv \ln {I(t+\Delta t)}- \ln {I(t)},
\label{eq:return}
\end{equation} 
where $I(t)$ is the market index at time $t$ and 
$\Delta t$ is the time-scale over which the fluctuation is observed. 
Market indices, rather than individual stock prices, have been
the focus of most previous studies as the former is more easily available,
and also gives overall information about the market. By contrast,
individual stocks are susceptible to sector-specific, as well as, stock-specific influences, 
and may
not be representative of the entire market. These two quantities, in fact,
characterize the market from different perspectives, the microscopic
description being based on individual stock price movements, while the 
macroscopic point of view focusses on the the collective market behavior
as measured by the market index.

\begin{figure*}
\includegraphics[width=0.95\linewidth]{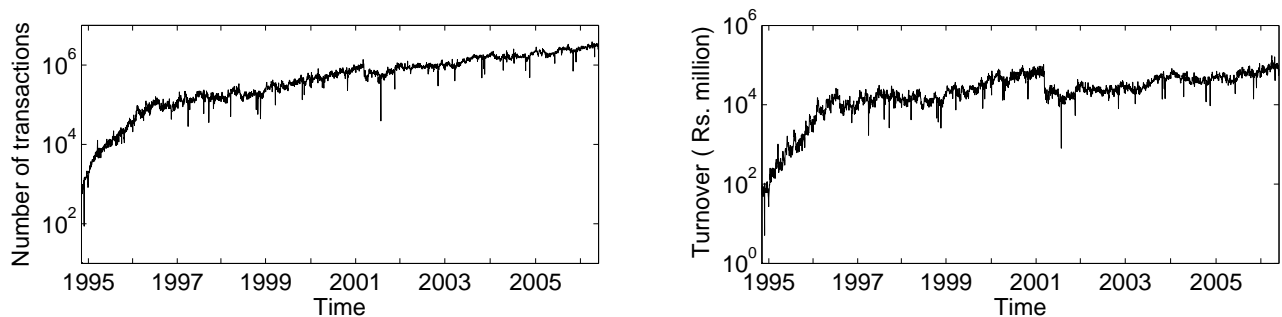}
\caption{Time evolution of the National Stock Exchange of India from
$1994-2006$ in terms of (left) the total number of trades and (right) the
total turnover (i.e., traded value).}
\label{fig:turnover}
\end{figure*}

The importance of interactions among stocks, relative to external
information, in governing market behavior has emerged only in recent times.
The earliest theories of market activity, e.g., Bachelier's random walk
model, assumed that price changes are the result of several independent
external shocks, and therefore, predicted the resulting distribution to be
Gaussian~\cite{Bachelier00}. As an additive random walk may lead to
negative stock prices, a better model would be a multiplicative random
walk, where the price changes are measured by logarithmic
returns~\cite{Osborne64}. While the return distribution calculated from
empirical data is indeed seen to be Gaussian at long time scales, at
shorter times the data show much larger fluctuations than would be expected
from this distribution~\cite{Fama65}. Such deviations were also
observed in commodity price returns, e.g., in Mandelbrot's analysis of
cotton price, which was found to follow a Levy-stable
distribution~\cite{Mandelbrot63}.  However, it contradicted the observation
that the distribution converged to a Gaussian at longer time scales. Later,
it was discovered that while the bulk of the return distribution
for the S\&P 500 index appears to be fit well by a Levy distribution, the
asymptotic behavior shows a much faster decay than expected. Hence, a
truncated Levy distribution, which has exponentially decaying tails, was
proposed as a model for the distribution of returns~\cite{Mantegna95}.
Subsequently, it was shown that the tails of the cumulative return
distribution for this index actually follow a power-law,
\begin{equation}
P_{c}(r > x) \sim x^{-\alpha}, 
\label{powerlaw}
\end{equation}
with the exponent $\alpha \approx 3$ (the ``inverse cubic
law'')~\cite{Gopikrishnan98}, well outside the stable Levy regime $0 <
\alpha < 2$.  This is consistent with the fact that at longer time scales
the distribution converges to a Gaussian.  Similar behavior has been
reported for the DAX, Nikkei and Hang-Seng
indices~\cite{Lux96,Gopikrishnan99}.  
These observations are somewhat surprising, although not at odds
with the ``efficient market hypothesis'' in economics,
which assumes that the
movements of financial prices are an immediate and unbiased reflection of
incoming news and future earning prospects. To explain these observations
various multi-agent models of financial markets have been proposed, where
the scaling laws seen in empirical data arise from interactions between
agents~\cite{Lux99}. Other microscopic models, where the agents (i.e., the
traders comprising the market) are represented by mutually interacting spins 
and the arrival of information by external fields, have also been used to 
simulate the financial market~\cite{Bornholdt01,Iori02,Chowdhury04,Sinha06b}. 
Among non-microscopic approaches, multi-fractal processes have
been used extensively for modelling such scale invariant
properties~\cite{Mandelbrot97,Bacry01}. The multi-fractal random walk model
has generalized the usual random walk model of financial price
changes and accounts for many of the observed empirical
properties~\cite{Bouchaud05}. 

However, on the empirical front, there is some controversy about the 
universality of the power-law
nature for the tails of the index return distribution. In the case of
developed markets, e.g., the All Ordinaries index of Australian stock
market, the negative tail has been reported to follow the inverse cubic law
while the positive tail is closer to Gaussian~\cite{Storer02}. Again, other
studies of the Hang Seng and Nikkei indices report the return distribution
to be exponential~\cite{Wang01,Kaizoji03}. For developing economies, the
situation is even less clear. There have been several claims that emergent
markets have return distribution that is significantly different from
developed markets. For example, a recent study contrasting the behavior of
indices from seven developed markets with the KOSPI index of the Korean
stock market found that while the former exhibit the inverse cubic law, the
latter follows an exponential distribution~\cite{Oh06}.  Another study of
the Korean stock market reported that the index distribution has changed to
exponential from a power-law nature only in recent years~\cite{Yang06}.  On
the other hand, the IBOVESPA index of the Sao Paulo stock market has been
claimed to follow a truncated Levy distribution~\cite{Miranda01,Gleria02}.
However, there have also been reports of the inverse cubic law for emerging
markets, e.g., for the Mexican stock market index IPC~\cite{Brizio05} and
the WIG20 index of the Polish stock market~\cite{Rak07}. A comparative
analysis of 27 indices from both mature and emerging markets found their
tail behavior to be similar~\cite{Jondeau99}.

Many of the studies reported above have only used graphical fitting to
determine the nature of the observed return distribution. This has
recently come under criticism as such methods often result in erroneous
conclusions. Hence, a more accurate study using reliable statistical techniques 
needs to be carried out to decide whether emerging markets do behave similar
to developed markets in terms of fluctuations. In this paper we have
carried out such a study for the Indian financial markets.
The Indian data is of unique importance in deciding whether emerging
markets behave differently from developed markets, as it is one of the
fastest growing financial markets in the world.  A recent study of
individual {\em stock prices} in the National Stock Exchange (NSE) of India
has claimed that the corresponding return distribution is exponentially
decaying at the tails~\cite{Matia04}, and not ``inverse cubic law'' that is
observed for developed markets~\cite{Lux96,Plerou99}. However, a more
detailed study over a larger data set has established the inverse cubic law
for individual {\em stock prices}~\cite{Pan07}.  On the other hand, to get
a sense of the nature of fluctuations for the entire market, one needs to
look at the corresponding distribution for the market index. 
Although the individual stock prices and the market index are related, it is
not obvious that they should have the same kind of distribution, as
this relation is dependent on the degree of correlation between different
stock price movements. While a
heavy-tailed distribution has been reported for the Nifty index of NSE, it
shows significant deviation from the inverse cubic law~\cite{Sarma05}.  In
this paper, we report analysis of {\em tick-by-tick} data for this index
along with a few others that fully characterizes the Indian market, to
conclusively establish the nature of their fluctuation distribution.

We focus on the two largest stock exchanges in India, the NSE and the
Bombay Stock Exchange (BSE). NSE, the more recent of the two, is not only
the most active stock exchange in India, but also the third largest in the
world in terms of transactions~\cite{ARS}.  We have studied the behavior of
this market over the entire period of its existence. During this period,
the NSE has grown by several orders of magnitude (Fig.~\ref{fig:turnover})
demonstrating its emerging character. In contrast, BSE is the oldest stock
exchange in Asia, and was the largest in India until the creation of NSE.
However, over the past decade its share of the Indian financial market has
fallen significantly. Therefore, we contrast two markets which have evolved
very differently in the period under study.  

We show that the Indian financial market, one of the largest emerging
markets in the world, has index fluctuations similar to that seen for
developed markets. Further, we find that the nature of the distribution is
invariant with respect to different market indices, as well as the
time-scale of observation. Taken together with our previous work on the
distribution of individual stock price returns in Indian
markets~\cite{Pan07,Sinha06}, this strongly argues in favor of the
universality of the nature of fluctuation distribution, regardless of the
stage of development of the market or the economy underlying it.

\section{Data Description}
Our primary data-set is that of the Nifty index of NSE which, along with
the Sensex of BSE, is one of the primary indicators of the Indian market.
It is composed of the top 50 highly liquid stocks which make up more than
half of the market capitalisation in India.  We have used (i)~high
frequency data from Jan 2003 -- Mar 2004, where the market index is
recorded every time a trade takes place for an index component. The total
number of records in this database is about $6.8 \times 10^{7}$. We have
also looked at data over much longer periods by considering daily closing
values of (ii)~the Nifty index for the 16-year period Jul 1990 -- May 2006
and (iii)~the Sensex index of BSE for the 15-year period Jan 1991 -- May
2006. In addition, we have also looked at the BSE 500 index for the much
shorter period Feb 1999 -- May 2006.  Sensex consists of the 30 largest and
most actively traded stocks, representative of various sectors of BSE,
while the BSE 500 is calculated using 500 stocks representing all 20 major
sectors of the economy.
\section{Distribution of Index Returns}
We first report the analysis of the high-frequency data for the NSE Nifty
index, which we sampled at 1-min intervals to generate the time series
$I(t)$.  From $I(t)$ we compute the logarithmic return $R_{\Delta t}(t)$, 
defined in Eq.~(\ref{eq:return}). These
return distributions calculated using different time intervals may have
varying width, owing to differences in their volatility, defined as
$\sigma^{2}_{\Delta t} \equiv \langle R^{2} \rangle-\langle R \rangle^{2}$, 
where $\langle \ldots \rangle$ denotes the time average over the given time
period.  Hence, to be able to compare the distributions, we need to
normalize the returns $R(t)$ by dividing them with  the volatility
$\sigma_{\Delta t}$. However, this leads to systematic underestimation of
the tail of the normalized return distribution. This is because, even when 
a single return $R(t)$ is very large, the scaled return is bounded 
by $\sqrt{N}$, as the same large return also contributes to the variance
$\sigma_{\Delta t}$. To avoid this, we remove the contribution of $R(t)$
itself from the volatility, and the new rescaled volatility is defined as 
\begin{equation}
\sigma_{\Delta t} (t)=\sqrt{\frac{1}{N-1} \sum_{t^\prime \neq t} \{R
(t^{\prime},\Delta t)\}^2
-\langle R (t^{\prime}, \Delta t) \rangle^2},
\end{equation}
as described in Ref.~\cite{Bouchaud03}. The resulting {\em normalized}
return is given by, 
\begin{equation}
r(t,\Delta t) \equiv \frac{R-\langle R \rangle}{\sigma_{\Delta t}(t)}.
\label{normalized_return}
\end{equation}

Prior to obtaining numerical estimates of the distribution parameters, 
we carry out a test for the nature of the return distribution, i.e., 
whether it follows a power-law
or an exponential or neither. For this purpose we use a statistical tool
that is independent of the quantitative value of the distribution parameters.
Usually, it is observed that the tail of the return distribution decays
at a slower rate than the bulk. Therefore, the determination of the
nature of the tail depends on the choice of the lower cut-off $u$ 
of the data used for fitting a theoretical distribution.
To observe this dependence on the cut-off $u$, we calculate the 
TP- and TE-statistics \cite{Pisarenko06,Pisarenko04} as a function of $u$,
comparing the behavior of the tail of the empirical distributions with 
power-law and exponential functional forms, respectively.
These statistics converge to zero if the underlying distribution
follows a power-law (TP) or exponential (TE), regardless of
the value of the exponent or the scale parameter (see Appendix \ref{app:tpte}).
On the other hand, they deviate from zero if the observed
return distribution differs from the target theoretical distribution
(power-law for TP and exponential for TE).

\begin{figure}
\includegraphics[width=0.9\linewidth]{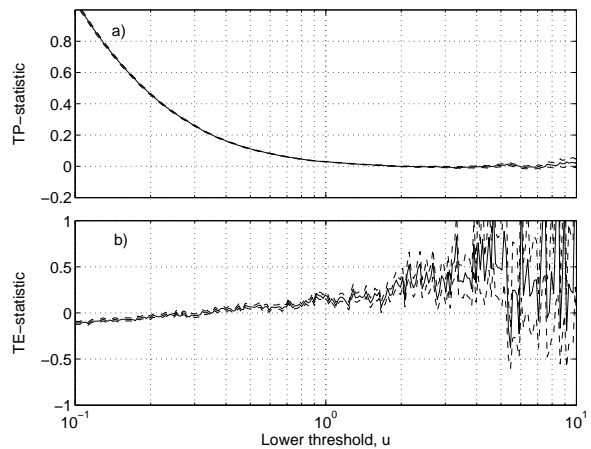}
\caption{(a) TP-statistic and (b) TE-statistic as function of 
the lower cut-off $u$ for positive returns of the NSE with time interval
$\Delta t$ = 1 min. The broken lines indicate plus or minus one standard
deviation of the statistics.} 
\label{fig:tpte}
\end{figure}
\begin{figure}
\includegraphics[width=0.8\linewidth]{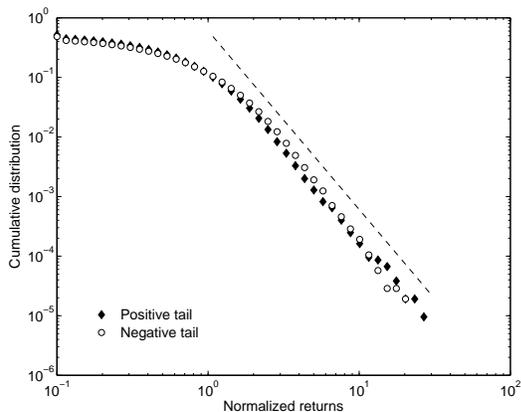}
\caption{The cumulative distribution of the normalized 1-min return for the
NSE Nifty index. The broken line indicates a power-law with exponent 
$\alpha = 3$.}
\label{fig:nifty_1min}
\end{figure}
Fig.~\ref{fig:tpte} shows visually the deviation of the empirical data
from the power-law and exponential distributions. The TP- and the TE-statistics 
are plotted as functions
of the lower cut-off $u$ for 1-min returns of the NSE Nifty index.
The TP-statistic shows a large deviation till $u\leq 1$, after which
it converges to zero indicating power-law behavior for large $u$. 
Correspondingly, the TE-statistic excludes an
exponential model for $u\geq 1$ as well as for very low values of $u$, 
although over the intermediate range $2\times 10^{-1}<u<6\times 10^{-1}$ 
an exponential approximation may be possible. 

Fig.~\ref{fig:nifty_1min} shows the cumulative distribution of the
normalized returns for $\Delta t = 1$ min.  For both positive and negative
tails, there is an asymptotic power-law behavior. The power-law regression
fit for the region $r \ge 2$ give exponents for the positive and the negative 
tails estimated as
\begin{equation}
\alpha = \left\{ \begin{array}{ll}
 2.98 \pm 0.09 & \mbox{(positive tail)} \\
 3.37 \pm 0.10 & \mbox{(negative tail).}
\end{array} \right.
\label{eqn:tail_exponent}
\end{equation}
Note that, to avoid artifacts due to data measurement process in the
calculation of return distribution for $\Delta t<1$ day, we have removed
the returns corresponding to overnight changes in the index value.

We also perform an alternative estimation of the tail index of the the
above distribution by using the Hill estimator~\cite{Hill75}, which is the
maximum likelihood estimator of $\alpha$. For finite samples, however, the
expected value of the Hill estimator is biased and depends crucially on the
choice of the number of order statistics used for calculation. We have used
the bootstrap procedure~\cite{Pictet98} to reduce this bias and to choose
the optimal number of order statistics for calculating the Hill estimator,
described in detail in the Appendix~\ref{app:hill}.  
We found $\alpha \simeq$~3.22 and
3.47 for the positive and the negative tails, respectively.

\begin{figure*}
\includegraphics[width=0.48\linewidth]{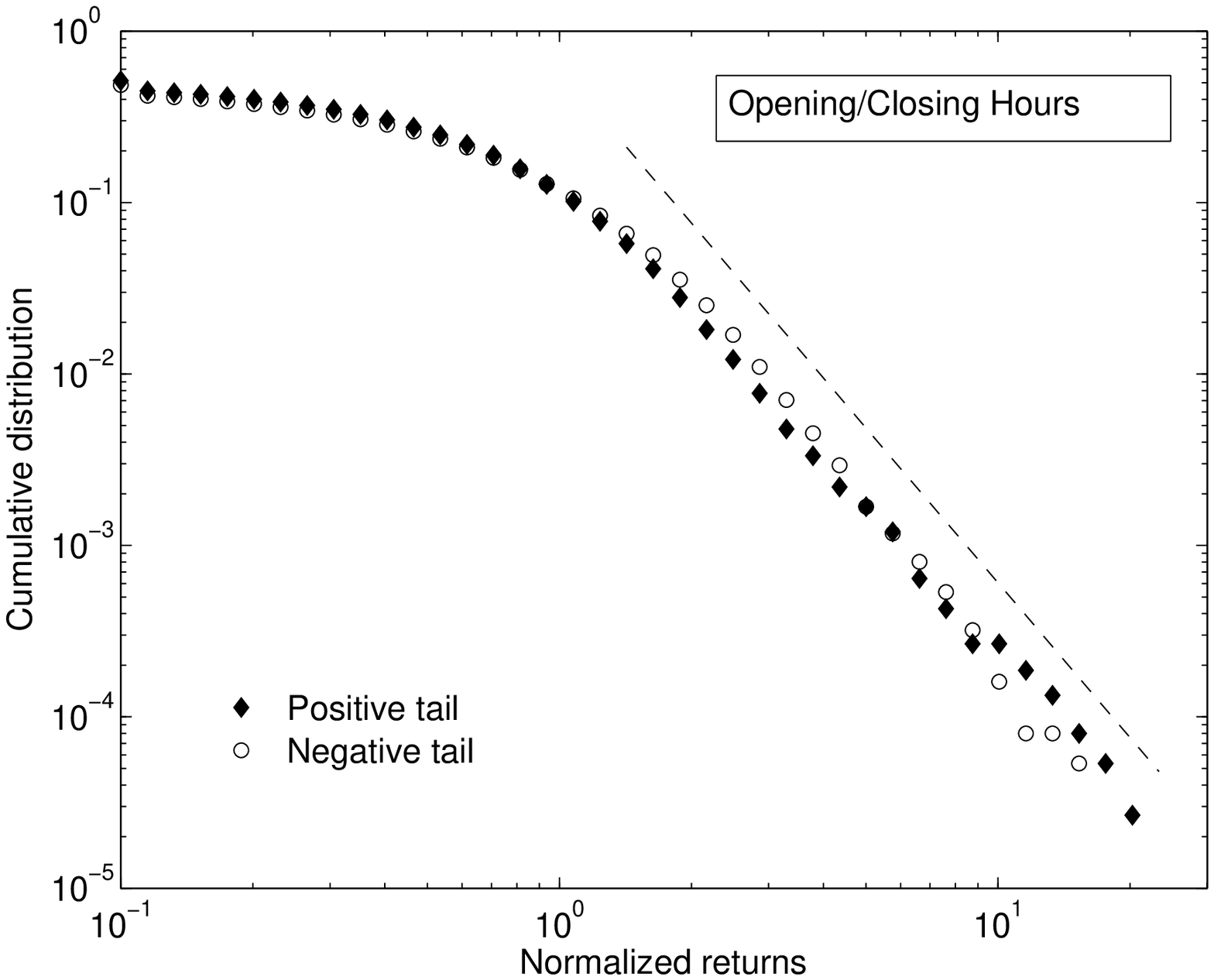}
\includegraphics[width=0.48\linewidth]{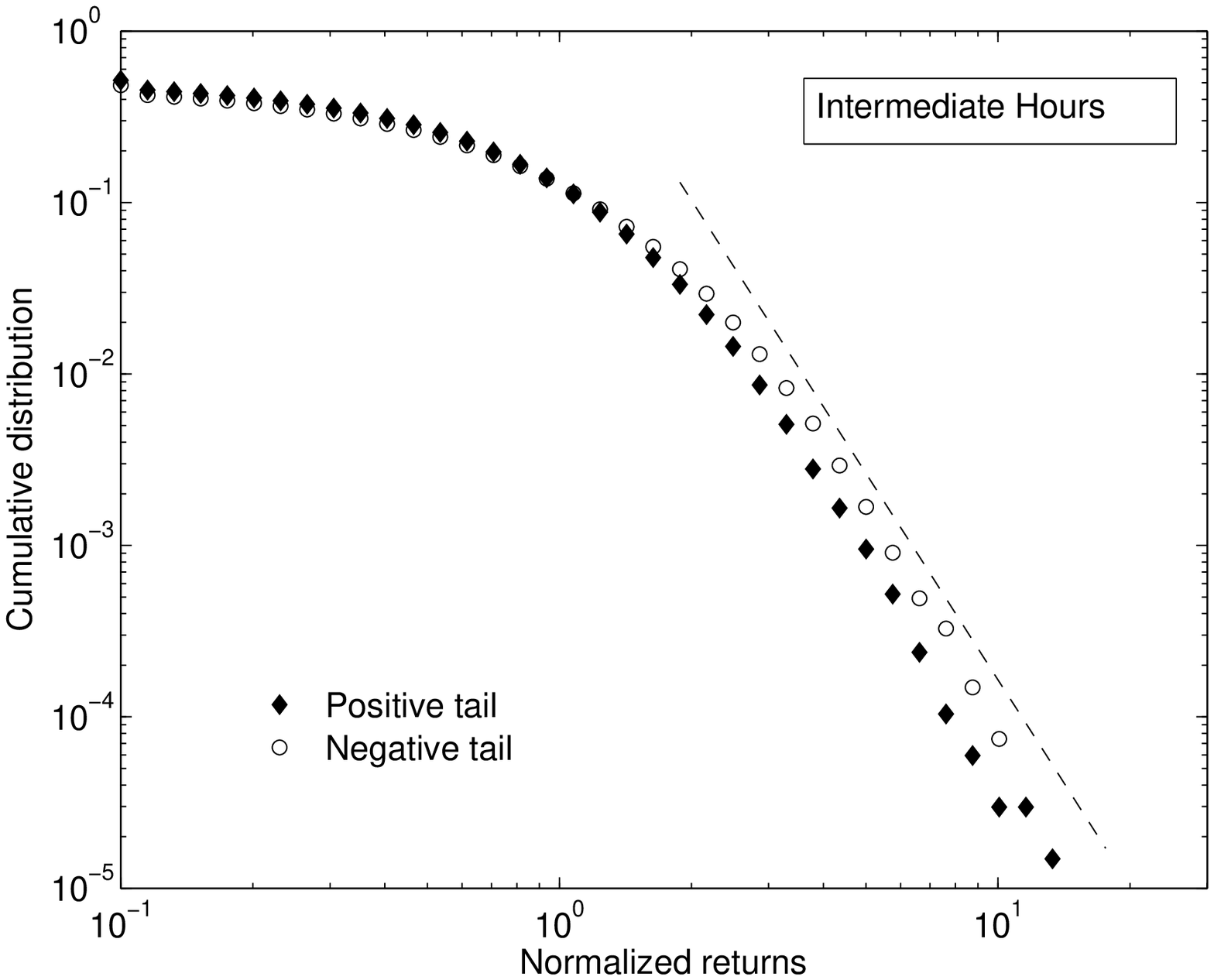}
\caption{
Intra-day variation in the cumulative distribution of the normalized 1-min
return for the NSE Nifty index: return distribution during (left) the
opening and closing hours (the broken line indicates a power law with
exponent $\alpha = 3$) and (right) the intermediate time period (the broken
line indicates a power law with exponent $\alpha = 4$).} 
\label{fig:open_clo}
\end{figure*}
To investigate the effect of {\em intra-day} variations in market activity,
we analyze the 1-min return time series by dividing it into two parts, one
corresponding to returns generated in the opening and the closing hours of
the market, and the other corresponding to the intermediate time period.
In general, it is known that the average intra-day volatility of stock returns
follows an U-shaped pattern~\cite{Wood85,Harris86} and one can expect this
to be reflected in the nature of the fluctuation distribution for the
opening and closing periods, as opposed to the intervening period.  We
indeed find the index fluctuations for these two data sets to be different
(Fig.~\ref{fig:open_clo}).  In particular, the cumulative distribution tail
for the opening and closing hour returns show a power-law scaling with
exponent close to $3$, whereas for the intermediate period we see that the
exponent is close to $4$. This observation is similar to that reported for
the German DAX index, where removal of the first few minutes of return data
after the daily opening resulted in a power-law distribution with a
different exponent compared to the intact data set~\cite{Lux01}.

\begin{figure}
\includegraphics[width=0.8\linewidth]{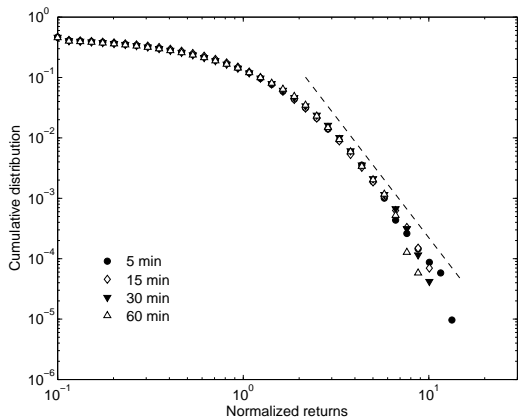}
\caption{The negative tail of the cumulative 
distribution of the NSE Nifty index returns for different time intervals
$\Delta t$ upto $60$ min. The broken line indicates a power-law with exponent 
$\alpha = 4$.}
\label{fig:nifty_comparison}
\end{figure}
Next, we extend our analysis for longer time scales, $\Delta t$. We find
that time aggregation of the data increases the $\alpha$ value.  The tail
of the return distribution still retains its power-law form
(Fig.~\ref{fig:nifty_comparison}), until at longer time scales the
distribution slowly converges to Gaussian behavior
(Table~\ref{tab:tail_exponent}).  The results are invariant with respect to
whether one calculates return using the sampled index value at the end
point of an interval or the average index value over the interval.
Figure~\ref{fig:nifty_comparison} shows the cumulative distribution of
normalized Nifty returns for time scales up to 60 min. However, using a
similar procedure for generating {\em daily} returns from the tick-by-tick data
would give us a very short time series. This is not enough for reliable
analysis as it takes at least 3000 data points for a meaningful estimate of
the tail index. 

\begin{figure}
\includegraphics[width=0.8\linewidth]{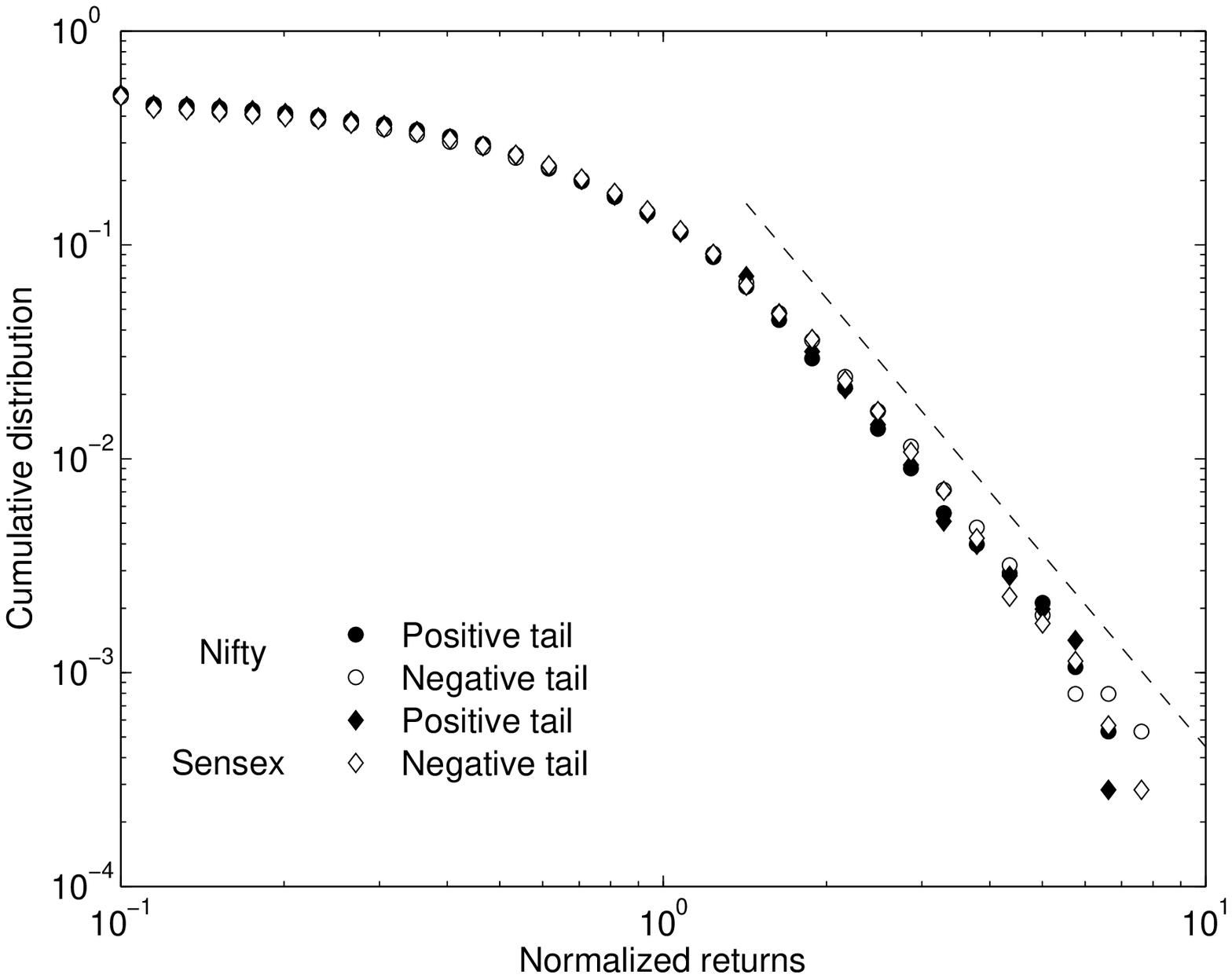}
\caption{The cumulative distribution of the normalized 1-day return for the
NSE Nifty and BSE Sensex index. The broken line indicates a power-law with
exponent  $\alpha = 3$.}
\label{fig:nifty_sensex}
\end{figure}

\begin{table*}[tbp]
\centering
\caption{Comparison of the power-law exponent $\alpha$ of the cumulative 
distribution function for various index returns. Power-law regression fits
are done in the region $r\ge2$. The Hill estimator is calculated using the
bootstrap algorithm.}
\label{tab:tail_exponent}       
\begin{tabular}{llllll}
\hline \noalign{\smallskip}
Index & $\Delta t$ &
\multicolumn{2}{c}{Power-law fit} & 
\multicolumn{2}{c}{Hill estimator}\\ \cline{3-6}\noalign{\smallskip}
 &  & \multicolumn{1}{c}{Positive} & \multicolumn{1}{c}{Negative} &
 \multicolumn{1}{c}{Positive} & \multicolumn{1}{c}{Negative} \\ 
\noalign{\smallskip}\hline\noalign{\smallskip}
Nifty ('03-'04) & 1 min & $2.98 \pm 0.09$ & $3.37 \pm 0.10$ & $3.22 \pm 0.03$ & $3.47 \pm
0.03$  \\
& 5 min & $4.42 \pm 0.37$ & $3.44 \pm 0.21$ & $4.51 \pm 0.03$ & $4.84 \pm
0.03$  \\
& 15 min & $5.58 \pm 0.88$ & $3.96 \pm 0.27$ & $6.25 \pm 0.03$ & $4.13 \pm
0.04$  \\
& 30 min & $5.13 \pm 0.41$ & $3.92 \pm 0.45$ & $5.65 \pm 0.03$ & $4.30 \pm
0.03$  \\
& 60 min & $5.99 \pm 1.52$ & $4.42 \pm 0.65$ & $7.85 \pm 0.03$ & $5.11 \pm
0.04$  \\
\noalign{\smallskip}
Nifty ('90-'06) & 1 day & $3.10 \pm 0.34$ & $3.18 \pm 0.28$ & $3.33 \pm 0.14$ & $3.37 \pm 0.14$  \\
\noalign{\smallskip}
Sensex ('91-'06) & 1 day & $3.33 \pm 0.77$ & $3.45 \pm 0.25$ & $2.93 \pm 0.15$ & $3.84 \pm
0.12$  \\ 
\noalign{\smallskip}\hline
\end{tabular}
\end{table*}
For this reason, we have analyzed the daily data using a different source,
with the time period stretching over a considerably longer period (16
years).  The return distribution of the daily closing data of Nifty shows
qualitatively similar behavior to the 1 min distribution. The Sensex index,
which is from another stock exchange, also follows a similar
distribution(Fig.~\ref{fig:nifty_sensex}). The measured exponent values are
all close to 3.  This does not contradict the earlier observation that
$\alpha$ increases with $\Delta t$, because, increasing the sample size (as
has been done for $\Delta t=1$~day) improves the estimation of $\alpha$.
This underlines the invariance of the nature of market fluctuations with
respect to time aggregation, interval used and different exchanges.

\section{Discussion and Conclusion}
The much shorter data-set of the BSE 500 daily returns shows a significant
departure from power-law behavior, essentially following an exponential
distribution (Figure not shown). This is not surprising, as looking at data
over shorter periods can result in misidentification of the nature of the
distribution. Specifically, the relatively low number of data points
corresponding to returns of large magnitude can lead to missing out the
long tail. In fact, even for individual stocks in developed markets,
although the tails follow a power-law, the bulk of the return distribution
is exponential~\cite{Silva04}.  This problem arising from using limited
data-sets might be one of the reasons why some studies have seen
significant deviation of index return distribution from a power-law.

A more serious problem is that the analysis in many of these studies is
usually performed only by graphically fitting the data with a theoretical
distribution function. Such a visual judgement of the goodness of fit
may lead to erroneous characterization of the nature of fluctuation
distribution. Graphical procedures are often subjective, particularly with
respect to the choice of the lower cut-off upto which fitting is carried out. 
This dependence of the theoretical 
distribution that best describes the tail on the cut-off,
has been explicitly demonstrated through the use of TP- and 
TE-statistic in this paper.
Moreover, recent studies have criticized the reliability of graphical
methods by showing that least square fitting for estimating
the power-law exponent tends to provide biased estimates, while the maximum
likelihood method produces more accurate and robust
estimates~\cite{Goldstein04,Brizio05a}. So we have used the Hill 
estimator to determine the tail exponents.

If the individual stocks follow the inverse cubic law, it would be
reasonable to suppose that the index, which is a weighted average of
several stocks, will also behave similarly, provided the different stocks
move in a correlated fashion~\cite{Gopikrishnan99}. As the price movements of stocks in an
emerging market are even more correlated than in developed
markets~\cite{Pan07a}, it is expected that the returns for stock prices
and the index should follow the same distribution. Therefore, the
demonstration of the inverse cubic law for the index fluctuations in the Indian 
market is consistent with our previous study \cite{Pan07} showing 
that the individual stock prices in this market follow the same behavior.

On the whole, our study points out the remarkable robustness of the nature
of the fluctuation distribution for Indian market indices.  While, in the
period under study, the NSE had begun operation and rapidly increased in
terms of activity, the BSE had existed for a long time prior to this period
and showed a significant decrease in market share.  However, both showed
very similar fluctuation behavior. This indicates that, at least in the
Indian context, the distribution of returns is invariant with respect to
markets.  The fact that the distribution is quantitatively same as
developed markets, implies that it is also probably independent of the
state of the economy.  In addition, our observation that the intra-day
return distribution of Indian market index show properties similar to that
reported for developed markets, suggest that even at this level of detail
the fluctuation behavior of the two kinds of markets are rather similar.
Therefore, our results indicate that although markets may differ from each
other in terms of (i)~the details of their components, (ii)~the nature of
interactions and (iii)~their susceptibility to news from outside the
market, there may be universal mechanisms responsible for generating market
fluctuations as indicated by the observation of invariant properties.
The rigorous demonstration of such a universal law for market behavior 
is significant for the physics of strongly interacting complex systems,
as it suggests the existence of robust features that are independent of 
individual details of different systems.

\section*{Acknowledgments}
We are grateful to M.~Krishna for invaluable assistance in obtaining and
analyzing the high-frequency NSE data. We thank the anonymous referee
for helpful suggestions.

\appendix

\section{TP-statistic and TE-statistic}
\label{app:tpte}
\subsection{TP-statistic}
Consider the power-law distribution,
\begin{equation}
F(x)=1-P_c(x)=1-(u/x)^{\alpha}, ~{\rm for}~ x \geq u,
\label{eq:pareto}
\end{equation}
where $u$ is the lower cut-off, and $\alpha$ is the power-law exponent for the
distribution. For a finite sample $x_1, \dots, x_n$, the TP-statistic, 
TP($x_1,\dots,x_n$), is defined such that it converges to zero asymptotically
for large $n$ \cite{Pisarenko06,Pisarenko04}. If the underlying
distribution for a sample differs from the power-law form given in 
Eq.(\ref{eq:pareto}), TP is seen to deviate from zero.
This statistic is based on
the first two normalized statistical log-moments of the power-law distribution,
\begin{equation}
E_{1}=E \left[\log\frac{X}{u}\right] =
\int_{u}^{\infty}\log\frac{x}{u}dF(x)=\frac{1}{\alpha}, ~{\rm and}
\label{eq:moment1}
\end{equation}
\begin{equation}
E_{2}=E \left[\log^2\frac{X}{u}\right] =
\int_{u}^{\infty}\log^2\frac{x}{u}dF(x)=\frac{2}{\alpha^2},
\label{eq:moment2}
\end{equation}
where, $E [z]$ represents the mathematical expectation of $z$.
The TP-statistic is then defined as
\begin{equation}
\mathrm{TP} =
\left[\frac{1}{n}\sum_{k=1}^{n}\log\frac{x_{k}}{u}\right]^2-\frac{1}{2n}\sum_{k=1}^{n}\log^2\frac{x_{k}}{u},
\label{eq:tpstat}
\end{equation}
which tends to zero as
$n \rightarrow \infty$. The estimation of the standard deviation for the TP
statistic is provided by the standard deviation of the sum
\begin{equation}
\left( \frac{E_{2}}{2}-2E_{1}^2 \right)+\frac{1}{n}\sum_{k=1}^{n}\left[2E_1
\log\frac{x_k}{u}-\frac{1}{2}\log^2\frac{x_k}{u}\right].
\label{eq:tpstd}
\end{equation}

\subsection{TE-Statistics}
Consider the exponential distribution,
\begin{equation}
F(x)=1-P_c(x)=1-\exp(-(x-u)/d),~{\rm for}~ x \geq u,
\label{eq:exponential}
\end{equation}
where $u$ is the lower cut-off, and $d (> 0)$ is the scale parameter
of the distribution. For a finite sample $x_1, \dots, x_n$, the TE statistic, 
TE($x_1,\dots,x_n$), is defined such that it converges to zero asymptotically
for large $n$ \cite{Pisarenko06}. If the underlying
distribution for a sample differs from the exponential form given in 
Eq.(\ref{eq:exponential}), TE is seen to deviate from zero.
This statistic is based on
the first two normalized statistical (shifted) log-moments of the 
exponential distribution,
\begin{eqnarray}
E_1 &=& E \left[ \log\left(\frac{X}{u}-1\right) \right] =
\int_{u}^{\infty}\log\left(\frac{x}{u}-1\right)dF(x) \nonumber \\
& = & \log\frac{d}{u}-\gamma,
\label{eq:moment3}
\end{eqnarray}
where $\gamma = 0.577215$ is the Euler constant, and
\begin{eqnarray}
E_2 &=& E \left[ \log^2\left(\frac{X}{u}-1\right) \right] =
\int_{u}^{\infty}\log^2\left(\frac{x}{u}-1\right)dF(x) \nonumber \\
&=& \left(\log\frac{d}{u}-\gamma\right)^2 + \frac{\pi^2}{6}.
\label{eq:moment4}
\end{eqnarray}
As before, $E[\dots]$ denotes the mathematical expectation.
The TE-statistic is then defined as 
\begin{equation}
\mathrm{TE} =
\frac{1}{n}\sum_{k=1}^{n}\log^2\left(\frac{x_{k}}{u}-1\right)
-\left[\frac{1}{n}\sum_{k=1}^{n}\log\left(\frac{x_{k}}{u}-1\right)\right]^2
- \frac{\pi^2}{6},
\label{eq:testat}
\end{equation}
which tends to zero as
$n\rightarrow \infty$. The estimation of the standard deviation for the 
TE-statistic is provided by the standard deviation of the sum
\begin{equation}
\frac{1}{n}\sum_{k=1}^{n}\left[\log\left(\frac{x_k}{u}-1\right)-E_1 \right]^2.
\label{eq:testd}
\end{equation}

\section{Hill estimation of the tail exponent $\alpha$}
\label{app:hill}

The Hill estimator gives consistent estimate of the tail exponent $\alpha$
from random samples of a distribution with an asymptotic power-law form.
For our analysis, we arrange the returns in decreasing order such that
$r_{1}>\cdots>r_{n}$.  Then the Hill estimator (based on the largest $k+1$
values) is given as
\begin{equation}
\gamma_{k,n}=\frac{1}{k}\sum_{i=1}^{k}{\log \frac{r_{i}}{r_{k+1}}},
\label{hill_estimator}
\end{equation}
for $k=1,\cdots,n-1$. The estimator $\gamma_{k,n} \rightarrow \alpha^{-1}$
when $k \rightarrow \infty$ and $k/n \rightarrow 0$.  However, for a finite
time series, the expectation value of the Hill estimator is biased, i.e.,
it will consistently over or underestimate $\alpha$.  Further, $\gamma$
depends critically on our choice of $k$, the order statistics used to
compute the Hill estimator. 

If the form of the distribution function from which the random sample is
chosen is known, then the bias and the stochastic error variance of the
Hill estimator can be calculated. From this, the optimum $k$ value can be
obtained such that the asymptotic mean square error of the Hill estimator
is minimized. Increasing $k$ reduces the variance because more data are
used, but increases the bias because the power-law is assumed to hold only
in the extreme tail. Unfortunately, the distribution for the empirical data
is not known and hence this procedure has to be replaced by an
asymptotically equivalent data driven process. 

One such method is subsample bootstrap method. This  method can be used to
estimate an optimal number for the order statistics ($\bar{k}$) that will
reduce the asymptotic mean square error  of the Hill estimator. However,
this process requires the choice of certain parameters, e.g., the subsample
size $n_s$ and the range of $k$ values in which one searches for the
minimum of the bootstrap statistics.  We briefly describe this procedure
below; for details and mathematical validation of this procedure, please
see Ref.~\cite{Pictet98}. 

We assume the underlying empirical distribution function to be 
heavy-tailed, viz., 
\begin{equation}
P_{c}(x)=ax^{-\alpha}[1+bx^{-\beta}+o(x^{-\beta})], 
\end{equation}
with $\alpha,\beta,a > 0$ and $-\infty <b<\infty$.  We first calculate an
initial $\gamma_{0}=\gamma_{k_{0},n}$ for the original series with a
reasonably chosen (but non-optimal) $k_0$. Then we choose various
subsamples of size $n_s$ randomly from the original series, which are
orders of magnitude smaller then $n$. The quantity $\gamma_{0}$ is a good
approximation of subsample $\alpha^{-1}$, since the error in $\gamma$ is
much larger for $n_s$ than for $n$ observations. The optimal order
statistics $\bar{k}_s$ for the subsample is found by computing
$\gamma(k_s,n_s)$ for different values of $k_s$ and then minimising the
deviation from $\gamma_{0}$. Given $\bar{k}_s$, the suitable full sample
$\bar{k}$ can be found by using 
\begin{equation}
\bar{k}=\bar{k}_s\left(\frac{n}{n_s}\right)^{\frac{2\beta}{2\beta+\alpha}}.
\end{equation}
Here the initial estimate of $\alpha$ is taken to be $1/\gamma_{0}$.
Further, we have considered $\beta=\alpha$, as done by Hall~\cite{Hall90},
although the results are not very sensitive to the choice of $\beta$.  Once
$\bar{k}$ is calculated, the final estimate of the tail index is given by
$\alpha=1/\gamma_{\bar{k},n}$.

For calculating the initial $\gamma_{0}$ we have chosen $k_{0}$ to be
$0.5\%$ of the sample size $n$. 1000 subsamples, each of size $n_s=n/40$,
are randomly picked from the full data set.  To obtain optimal $k_s$, we
confine ourselves to $4\%$ of the subsample size $n_s$.  To find the
stochastic error in our estimation of $\alpha$, we have computed the 95\%
confidence interval as given by $\pm 1.96[1/(\alpha^2m)]^{1/2}$.  Although
a Jackknife algorithm can also be used to calculate this error bound, the
results obtained using this method will be close to that obtained using the
bootstrap method over many realizations~\cite{Pictet98}, as we have done in
this paper.

\bibliography{referenc}
\end{document}